\documentclass[apjl]{emulateapj}

\def\kms{\rm\,km\;s^{-1}}

\begin{document}
\title{Galactic-center S~Stars as a prospective test of the \\
Einstein Equivalence Principle}
\author{Raymond Ang\'elil}
\author{Prasenjit Saha}
\affil{Institute for Theoretical Physics, University of Z\"urich, \\
  Winterthurerstrasse 190, CH-8057 Z\"urich, Switzerland }
\date{\today}
\begin{abstract}
  The S~Stars in the Galactic-center region are found to be on
  near-perfect Keplerian orbits around presumably a supermassive black
  hole, with periods of 15--50 yr.  Since these stars reach a few
  percent of light speed at pericenter, various relativistic effects
  are expected, and have been discussed in the literature.  We argue
  that an elegant test of the Einstein equivalence principle should be
  possible with existing instruments, through spectroscopic monitoring
  of an S~star concentrated during the months around pericenter,
  supplemented with an already-adequate astrometric determination of
  the inclination.  In essence, the spectrum of an S~star can be
  considered a heterogeneous ensemble of clocks in a freely-falling
  frame, which near pericenter is moving at relativistic speeds.
\end{abstract}


\maketitle

\section{Introduction}

The equivalence of gravitational and inertial mass, or that gravity
can be cancelled by transforming to a freely-falling frame, was tested
within classical celestial mechanics to high precision by the end of
the nineteenth century.  After all, if Mercury had a gravitational
constant differing from (say) Jupiter's by one part per million,
Mercury's orbit would not have fitted classical dynamics well enough
to highlight the anomalous precession ($43''$ per century, or
$10^{-7}$) that was later explained by general relativity.

The Einstein equivalence principle (EEP) adds to the classical or weak
equivalence principle the further physical postulate that special
relativity holds locally in a freely-falling frame.  A consequence of
the EEP is that the effects of gravity on test particles are fully
described by endowing spacetime with a Riemannian metric, and having
the particles follow geodesics of that metric. A further consequence
of the EEP is that the temporal metric component is, to leading order,
given by $g_{tt}= 1 + 2\Phi$, where $\Phi$ is the Newtonian
potential. Hence gravity causes time dilation.

Different aspects of the EEP have been verified by multiple
experiments, as reviewed extensively by \cite{will}. In
particular, null-redshift experiments test that gravity ---whatever it
does--- does the same to clocks of different types. Pound-Rebka
experiments verify that, in a static gravitational field, time is
dilated by a factor $g_{tt}^{-1/2} \simeq 1-\Phi$. GPS satellites and
binary pulsars are effectively freely-falling clocks moving
at\footnote{We use geometrized units: $GM=c=1$, where $M$ is the
  black-hole mass.} $\sim10^{-5}$ and $\sim10^{-3}$ respectively, and
exhibit time dilation of $g_{tt}^{-1}\simeq 1-2\Phi$.

But excellent as the existing experiments on the EEP are, it would be
even nicer to have a laboratory with experiments on multiple
materials, the whole freely-falling at relativistic speeds.  In this
Letter, we suggest that the S~stars in the Galactic-center region
approximate such a laboratory. Stellar spectra contain absorption
lines from different atomic species, which can be regarded as
different clocks.  The EEP asserts that special relativity holds in
the star's frame for all atomic processes (Local Lorentz Invariance),
irrespective of where the star is (Local Position Invariance), and
these can be tested by observations of a single star.  The remaining
ingredient of the EEP, namely that a freely falling frame does not
itself depend on the composition of the star (Universality of Free
Fall) requires multiple stars. The nature of a violation of the
universality of free fall is difficult to speculate on, however one
could imagine for example, fits from different S~Stars' redshift
curves all yielding different values for the black hole mass. Such a
deduction would signal a violation of the UFF.

Going beyond the EEP, yet another possibility is that a star with a
relativistically significant gravitational self energy, i.e., a pulsar,
may give different results.  Such a result would be a violation of the
Strong Equivalence Principle. It is expected that a significant number
of pulsars inhabit the inner milliparsec, and although none have yet
been found, searches are currently underway\citep{marquart}. 

The S~stars achieve the highest speeds of any known geodesics. For
example, S2 reaches $v > 0.03$ at pericenter, which it last passed in
2002 and will again in 2018.  The pericenter speed is an order of
magnitude faster than for known binary pulsars.  The spatial scales at
pericenter are of order a light day.  Hence, even if the quantitative
constraints are initially modest, the S~stars would test the EEP at
velocity and spatial scales not reached by other experiments.

Astrometric and redshift observations of the S~stars show them to be
on orbits (so far) indistinguishable from pure
Keplerian \citep{VLTandKECK, ghez, gillessen}.  Keplerian elements are
known to high accuracy for many stars.  For example, for S2 the
orbital inclination is measured to be $135\pm1^\circ$.  Expected
relativistic effects have been discussed extensively in the
literature.  These include not only time
dilation \citep{gillessenredshift}, but also secular orbit precession
\citep{rubilar,will2}
plus Newtonian perturbations from other stars \citep{merritt},
kinematic effects due to space curvature and
frame-dragging \citep{kannan,pretosaha}, and composite redshift
perturbations including light-path effects \citep{paper1,paper2}.

The EEP implies a time dilation of $1+2/r$. Hence a spectral line
intrinsically at $\nu_0$ will be redshifted to $\nu$, where
\begin{equation}\label{crux}
\frac{\nu_0}\nu = \left(1 + v_{\rm los}\right) \left(1+\frac2r\right) .
\end{equation}
Redshift is conventionally defined as $\nu_0/\nu-1$.  It is however,
somewhat more convenient if one defines $\ln(\nu_0/\nu)$ as the
redshift, and we will do so.  In any case, the $2/r$ term constitutes
a redshift perturbation of $\mathcal{O}\left(v^2\right)$ and is the
strongest relativistic perturbation. In other words, the leading-order
perturbation due to relativity is time dilation on a Keplerian
orbit. Space curvature perturbs the redshift at
$\mathcal{O}\left(v^3\right)$, frame dragging due to black-hole spin
at $\mathcal{O}\left(v^4\right)$, with additional physical effects
continuing at higher orders \citep{paper2}.  One way to test the EEP
would be to replace $2/r$ in Equation~(\ref{crux}) by $\alpha/r$ and
fit observed redshift curves for the parameter $\alpha$.  In fact,
this has already been attempted \citep{gillessenredshift} but the
available data appeared not yet sufficiently accurate to overcome
systematic uncertainties.  We argue below that observations
concentrated near pericenter will be particularly useful.

Relativistic perturbations to the S~stars increase with decreasing
pericenter distance. In orbit fitting from spectroscopy, as one looks
to stars with longer periods, the perturbations to the redshift from
relativity become overwhelmed by perturbations due to massive
perturbers. Of all the compact objects orbiting the central black
hole, we see only the brightest stars. Thousands of solar masses worth
of compact objects and a possibly significant dark matter component is
expected to inhabit the inner arcseconds. Our lack of understanding of
the form of this extended system may play a significant role in
obscuring relativistic perturbations for large orbits.  However,
current models for the Newtonian perturbations are estimated to most
likely be weaker than the time dilation effect for the shortest-period
stars such as S2\citep{schodel, merritt}. Hence we will disregard the the extended-mass
component in our analysis.

\section{\label{sec:level2} Observables}

Consider a star on a pure Keplerian orbit.  Elementary celestial
mechanics gives the position of the star parametrically, that is, both
coordinates and time are expressed as functions of the so-called
eccentric anomaly $\psi$.  At time
\begin{equation}\label{emitted_time}
t = \frac{P}{2\pi}\left(\psi-e \sin\psi\right)
\end{equation}
the position of the star in its orbital plane is
\begin{equation}\label{inplane}
(x,y,z) = \frac{P}{2\pi\sqrt a}
\left( \cos\psi - e, \sqrt{1-e^2}\sin\psi, 0 \right) .
\end{equation}
Here $P$ is the period, $e$ is the eccentricity, while $a$ is the
semimajor axis in units of the gravitational radius of the black hole.

We now rotate the coordinate system, first by the argument of
periapsis $\omega$ about the $z$~axis, then by the inclination $I$
about the new $x$~axis.  The observer is now along the new
$z$~axis. The redshift is 
\begin{equation}\label{final_redshift}
\ln \frac{\nu_0}\nu = \frac{dz}{dt} + \frac{\alpha}r =
\frac{A_C \, f(e,\omega,\psi) + A_R}{1-e \cos\psi},
\end{equation}
where
\begin{equation}
f(e,\omega,\psi) \equiv
\sqrt{1-e^2} \, \cos\omega \, \cos\psi - \sin\omega \, \sin\psi
\end{equation}
and we have introduced the coefficients
\begin{equation}\label{funky_parameters}
A_C\equiv \frac{\sin I}{\sqrt{a}} \quad \textrm{and} \quad
A_R\equiv \frac{\alpha}{a}.
\end{equation}
which can be interpreted as the amplitudes of the classical and
relativistic contributions to the redshift.  We also note that $t$
above is the coordinate time of emission. The time of observation is
of course $t$ plus the light travel time, whose varying part is the
R\o mer time delay, which is the $z$ coordinate of the star
\begin{equation}
A_C \frac{P}{2\pi}
\left[\left(\cos\psi-e\right) \sin\omega +
      \sqrt{1-e^2}\sin\psi  \cos\omega \right],
\end{equation}
and can also be interpreted as the time-integral of the redshift.

Classically, the redshift curve determines only the combination $\sin
I/\sqrt a$, leaving the inclination unknown.  Hence, for a given
redshift curve, the inferred orbital speed becomes infinite if the
orbit is face-on.  Not surprisingly, relativity prevents that
happening. From (\ref{final_redshift}) and (\ref{funky_parameters})
it follows that
\begin{equation}\label{strategy}
\sin^2I = \frac{A_C^2}{A_R}\alpha .
\end{equation}
Hence time-dilation breaks the inclination degeneracy if $\alpha$ is
known, or allows $\alpha$ to be measured if $I$ is known from astrometry. 

It is worth mentioning that another way in which relativity can break
the inclination degeneracy is through the well-known precession of
\begin{equation}
\Delta\omega = \frac{6\pi}{a(1-e^2)}
\end{equation}
per orbit.  For binary pulsars, the cumulative $\omega$ precession
amounts to several degrees per year, allowing $a$ and hence $I$ to be
inferred \citep{1975PAZh....1....5B}.  For Galactic-center stars,
however, the $\omega$ precession builds up much more slowly. Over a
single orbit the redshift due to precession is
$\sim\mathcal{O}\left(v^3\right)$ because $\Delta\omega\sim
\mathcal{O}\left(v^2\right)$ over one orbit.

\section{Parameter recovery}
We treat the following seven parameters as unknown.
 \begin{enumerate}
 \item The period $P$.
 \item An additive constant on $t$.
 \item The eccentricity $e$.
 \item The argument of pericenter $\omega$, which along with the
   inclination $I$ sets the orbit orientation with respect to the
   observer. Since the observer is on the $z$~axis, the nodal
   angle $\Omega$ corresponds to a rotation around the line of sight,
   which leaves the redshift curve invariant.
 \item The intrinsic frequency $\nu_0$.  This is the absolute line
   calibration, plus any offset due to the observer's radial
   velocity. Changing $\nu_0$ shifts the redshift curve vertically.
   The tangential motion of the observer is neglected here, since the
   Sun's orbital speed of $\sim200\kms$ in the Galaxy would contribute
   no significant Doppler shift.
\item A classical redshift amplitude $A_C$.
\item A relativistic redshift amplitude $A_R$. In fact, $a,I,\alpha$
  form a degenerate trio and are absorbed into the non-degenerate
  dummy parameters $A_C$ and $A_R$ via (\ref{funky_parameters}).
\end{enumerate}
Astrometry independently measures the first four of these, as well as
$I$.  Astrometric observations also involve six other parameters: the
position and proper motion of the Galactic center on the sky, the
distance to the Galactic center, and $\Omega$.  The mass of the black
hole is not an independent parameter, since it is a function of $P$
and $a$.  For our purposes, only $I$ from astrometry is essential. In
testing the EEP, while we shall argue that observation of the star
over a short time around pericenter suffices for spectroscopy,
recovery of the inclination via astrometry can be done anywhere on the
orbit.

It is important to note $\nu_0$ and $\nu$ need not refer to a single
spectral line.  The spectra of S~stars \citep[see e.g.,][]{martins}
contain multiple features.  Most are early-type stars with H and He
features, while about 10\% are late-type stars with molecular and
metal bands/lines and little or no H or He.  Hence, S-star spectra
could, in principle, test the equivalence principle for multiple
atomic processes.  If the stellar atmosphere does not change
appreciably over an orbit, an observed spectrum can be
cross-correlated on a logarithmic wavelength scale with a spectrum
observed at some other epoch, and the cross-correlation peak would
directly give the redshift $\ln(\nu_0/\nu)$ with $\nu_0$ an unknown
constant. If different atomic/molecular species behave differently in
a freely-falling frame, the shape of the cross-correlation curve would
change.  Alternatively, multiple spectral features could be fitted
simultaneously with variable redshift.  We do not, however, attempt to
model the observable spectra explicitly in this paper.

\begin{figure}[!t]
  \includegraphics[scale=0.6]{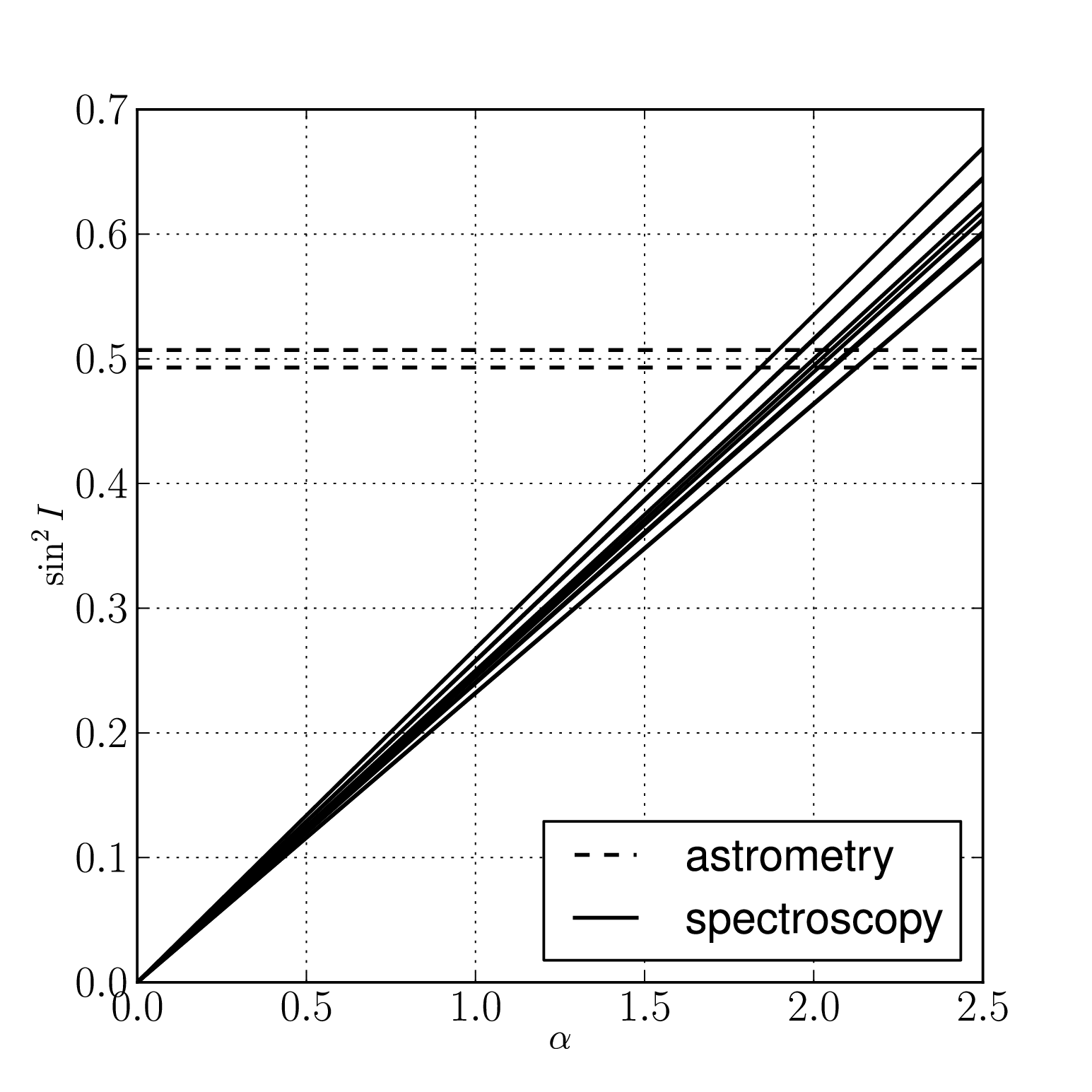}
  \caption{Recovery of $\alpha$ for ten mock data sets.  The ratio
    $A_C^2/A_R$ is the slope of the lines in the above plot, and is
    recovered from mock spectroscopic data of 14 data points with
    accuracy $10\kms$, concentrated around pericenter. The horizontal
    lines are the upper and lower confidence levels for the recovered
    inclination from astrometry, taken from \cite{gillessen}. The
    intersection point corresponds to the value of $\alpha$ for which
    both datatypes agree on the inclination.}\label{fig:getting_alpha}
 \end{figure}

\begin{figure}[!t]
  \includegraphics[scale=0.6]{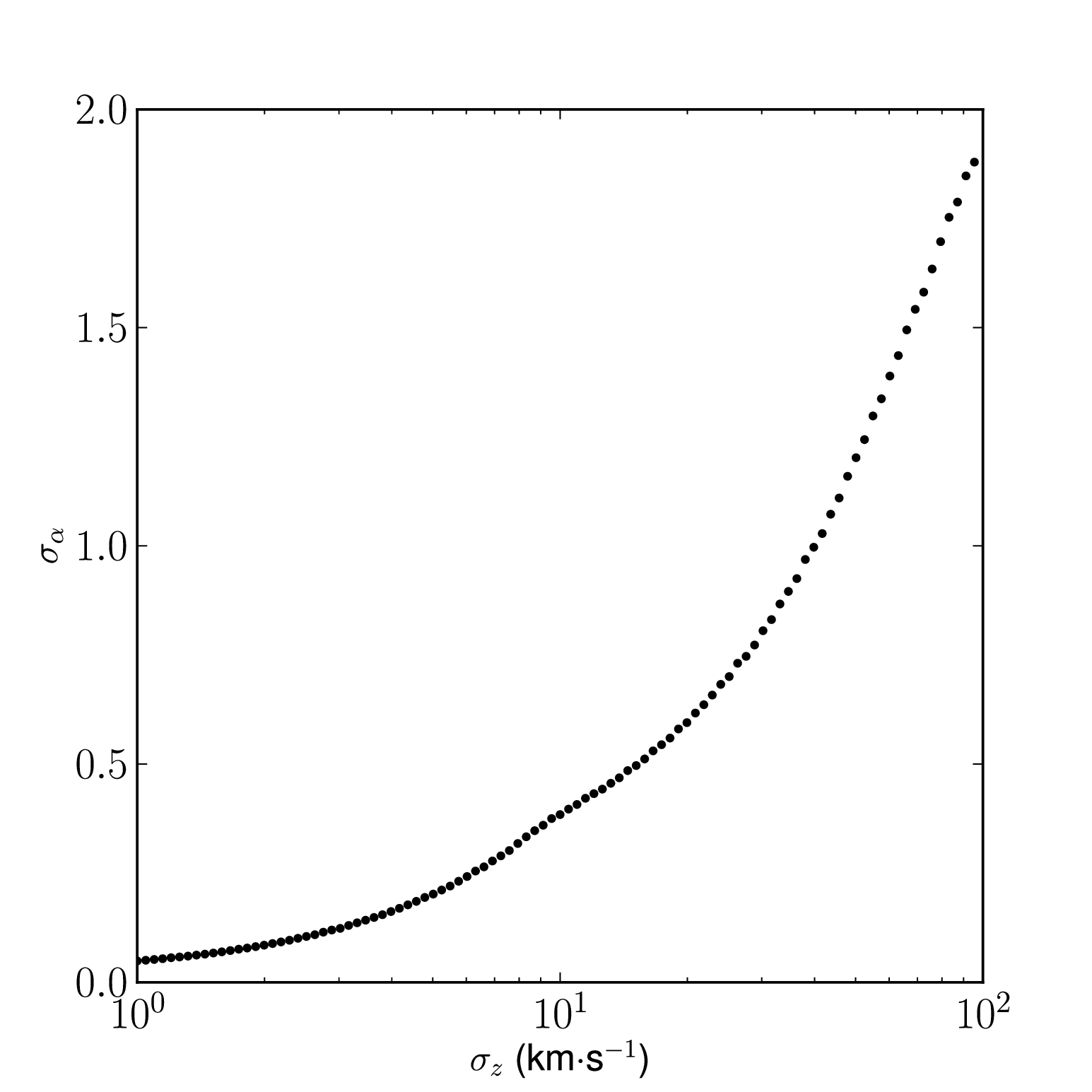}
  \caption{The $1\sigma$ level in the recovered value of $\alpha$ for
    spectroscopic data of different accuracy. At each data accuracy
    level, we have performed the same procedure illustrated in Figure
    \ref{fig:getting_alpha}, except that 60,000 mock data realizations
    have been used.}\label{fig:recovering_alpha}
\end{figure}

We now simulate the recovery of $\alpha$ as follows. We generate 10
mock redshift data points of S2 taken over two months at pericenter,
plus four additional data points, at $\pm 1,\pm2$yr around
pericenter. The data are generated with $\alpha = 2$, and orbital
parameters taken from \cite{gillessen}. To them we add gaussian random
noise at a dispersion of $10\kms$, and then fit via the seven
parameters.  We then assume $I$ has been measured by astrometry and
use (\ref{strategy}) to recover $\alpha$. Figure
(\ref{fig:getting_alpha}) shows an example for a few mock data
realizations at a fixed accuracy, and figure
(\ref{fig:recovering_alpha}) shows the dependency of the recovered
value of alpha with the data accuracy.

\section{Discussion}

Testing the equivalence principle using a combination of spectroscopy
and astrometry seems possible in the near future. In comparing the
spread in $A_C^2/A_R$ from mock data to the recovered value for $I$
from real astrometric data (illustrated in
Fig. \ref{fig:getting_alpha}), in testing the Equivalence Principle
using S2, the current accuracy available from astrometry sits at a
comfortable level. Spectroscopic accuracy of S2 at $10\kms$ is not yet
available, but seems plausible with future observations.  For the
late-type star S35, which has a more favorable spectrum, a fit error
of $10\kms$ has been achieved \citep{gillessen}.  We remark that any
systematic errors that do not change between observations are
harmlessly absorbed into $\nu_0$.

Naturally, in detecting relativistic effects on S~Stars, data during
pericenter passage are of greatest value. With instrumentation
currently available, an observation program concentrated over two
months during S2's next pericenter passage (2018) will prove to be
sufficient as a test for the Einstein Equivalence Principle. Figure
(\ref{fig:recovering_alpha}) argues that a small handful of spectral
measurements of S2 at $10\kms$ around pericenter imply a $1\sigma$
accuracy on $\alpha$ of $\sim 0.3$.

The approach we have taken above focuses on the essentials. The
degeneracy between $\alpha$, $I$ and $a$ for spectroscopy has been
lifted by using astrometry only to provide $I$. In practice, all the
parameters are fitted simultaneously to both astrometry and
spectroscopy. We have done simulations to verify that when this is
done, the degeneracy is implicitly broken by the mechanism highlighted
in this Letter.

Relativistic effects can be expected to become increasingly important
as corrections in other astrophysics relating to the S~stars.  Three
areas where this can be expected are the following.

\begin{enumerate}
\item The combination of spectroscopic and astrometric S~Star data
  provides us with the distance to the center of the
  galaxy. Astrometry is sensitive to the angular size of the orbit,
  while spectroscopy on the physical size. The quotient is the
  distance to the galactic center \citep{geometric}.
\item The position of the observed line depends on the velocity of the
  BH-star system with respect to the Earth. Spectroscopy therefore has
  the power to determine our velocity with respect to the central
  black hole \citep{paper2}, thus constraining the $U$ component of
  the Galactic local standard of rest.
\item The form of the mass distribution within the inner arcsecond
  affects the S~Star orbits. A better understanding of the density
  profile will provide insight into the region's star capture and
  formation history, and to the central dark matter distribution
  \citep{ghez,gillessen}.
\end{enumerate}   

In exploring 1 and 2, one cannot easily avoid relativity simply by
considering stars with larger orbit sizes (3), as perturbations due to the
enclosed mass become a problem. With the accuracy regime that
spectroscopy will enter in the the coming decade, one of two types of
perturbations to the redshift must be faced: either those from the
extended mass distribution, for S~Stars with large orbits, or, those
perturbations from relativity, for smaller orbit S~Stars. While
effects due to relativity are well understood in principle, and can be
easily treated, the constitution of the extended system is poorly
understood, and expects a more grueling treatment. In the coming
decade, spectroscopic S~Star accuracy at $\sim 10\kms$ is expected to
be available, and the discovery of stars closer in to the black hole
is anticipated. In light of these prospects, so that constraints on
these quantities may be improved, relativistic perturbations, while
interesting in their own right, can no longer be ignored and must be
faced.

\acknowledgements The authors thank S. Gillessen, G. F. R. Ellis, and
J.-P. Uzan for discussion and comments.

\end{document}